\documentclass[12pt,nofootinbib]{revtex4-2}
\usepackage{amsmath,amsfonts}
\usepackage{graphicx}
\usepackage{qcircuit}
\newcommand\ket[1]{|#1\rangle}
\newcommand\vev[1]{\langle #1\rangle}
\newcommand\bra[1]{\langle #1|}

\begin{document}
\title{Exotic equilibration dynamics on a 1-D quantum CNOT  gate lattice.}
\author{David Berenstein, Jiayao Zhao}

\address{Department of Physics, University of California at Santa Barbara, Santa Barbara, CA 93106 }

\begin{abstract}
We consider the dynamics of local entropy and nearest neighbor mutual information of a 1-D lattice of qubits via the repeated application of nearest neighbor CNOT quantum gates. This is a quantum version of a cellular automaton.
We analyze the entropy dynamics for different initial product states, both for open boundary conditions, periodic boundary conditions and we also consider the infinite lattice thermodynamic limit.
The  dynamics gives rise to fractal behavior, where we see the appearance of the Sierpinski triangle both for states in the computational basis and for operator dynamics in the Heisenberg picture.
In the thermodynamics limit, we see equilibration with a time dependence controlled by $\exp(-\alpha t^{h-1})$ where $h$ is the fractal dimension of the
Sierpinski triangle, and $\alpha$ depends on the details of the initial state. 
We also see log-periodic reductions in the one qubit entropy where the approach to equilibrium is only power law.
For open boundary conditions we see time periodic oscillations near the boundary, associated to subalgebras of operators localized near the boundary that are mapped to themselves by the dynamics.  

\end{abstract}

\maketitle
\section{Introduction}

The dynamics of non-equilibrium systems is a central theme of many-body dynamical systems.
The equilibration and thermaization dynamics of isolated quantum systems is not yet well understood \cite{cazalilla2010focus}. 
It is thus important to find systems where the dynamics of equilibration can be well understood analytically ab initio.

In this paper we are interested in studying one example of a system with translation invariance on a lattice, with the possibility of 
different choices of boundary conditions. 
Such systems are theoretical laboratories to analyze possible behaviors for quantum field theory equilibration phenomena.

 Considering the recent advances in quantum control and digital quantum computing devices, see for example \cite{arute2019quantum}, one can imagine situations where dynamical systems of many 
qubits in various dimensions, mimicking field theories, can be replicated and studied directly on a quantum device. In that case, the putative systems where the dynamics of equilibration is explored efficiently (from the theoretical point of view) will need to satisfy additional criteria, so that they can be realized realistically in one of these Noisy Intermediate-Scale Quantum (NISQ) devices. First, the dynamics should result from composition of quantum gates
between  pairs of qubits. Ideally, these will be physically near each other in the quantum device.
Secondly, the dynamics must be discrete in time, rather than continuous time. 
Finally, the initial state in the device should be simple. For example, a product state over individual qubits. 
These conditions are there to ensure that calculations on the quantum device do not take too long and that they might be carried out without having to implement  quantum error correction to get to the final answer.
In this case, if we have an efficient way to understand the dynamics from the theory, we can ask to what extent the quantum device reproduces the theory. To the extent that the quantum simulation  on a quantum computer can be considered as the system itself, the simple system we are studying is realized in nature. In this case, the failure of the NISQ device to reproduce the theory can be used to characterize the device instead. One can then  explore possible error mitigation measures one can take, which will probably depend on the device where the dynamics is implemented.
Here, we wish to report on an interesting such setup, where the dynamics of equilibration can be studied for initial product states, and where the evolution of the system can be computed classically very efficiently.

The system is described by the following generic 1-D dynamics, where the $U$ gate is applied alternately between nearest neighbor pairs:
\begin{equation}
\Qcircuit @C=1em @R=.7em {
& \qw &\multigate{1}{U} &\qw &\qw\\
& \qw &\ghost{U} &\multigate{1}{U}&\qw\\
&\qw &\multigate{1}{U} & \ghost{U}& \qw\\
&\qw &\ghost{U}& \qw&\qw
}\label{eq:qcircuitexample}
\end{equation}
This setup has the advantage that in principle it is easily parallelizable, it has a discrete time translation invariance
(every two layers of gates), and a discrete translation invariance by two steps, except at the boundaries.
If an additional $U$ interaction is added between the first and last qubit we can implement periodic boundary conditions. On the other
hand, if nothing is done at the edges we say we have open boundary conditions.
Our goal is to study this particular dynamics with both possible sets of boundary conditions where the gate $U$ is the CNOT quantum gate.
Also, we choose our initial state to be given by a product of state $\otimes^n \ket \psi$.

This dynamics has been studied before in \cite{gopalakrishnan2018facilitated} from the point of view of eigenstate thermalization hypothesis \cite{deutsch1991quantum,PhysRevE.50.888}
and in \cite{Berenstein:2018zif} as a limit of quantum chaotic dynamical systems related to generalized cat map actions (see also \cite{Berenstein:2015yxu}).
We should note that in this last paper it was noted that the system had strange properties for entanglement entropy evolution depending on the properties of the initial state for a special class of states and in the case of open boundary conditions. This also bears some similarity with the system studied in \cite{alba2019operator}. In this paper our goal is to understand what is responsible for such behavior.

The system is simple to analyze because it belongs to a general class of systems that is easy to simulate classically \cite{Gottesman:1998hu} by using the Heisenberg representation of the dynamics. 
It also bears close resemblance to Clifford cellular automata. These have been classified \cite{schlingemann2008structure,gutschow2010time} and differ from the above in one crucial respect: they are translation invariant with period one rather than two.
This has important consequences. The classification of these cellular automata produce a symmetric evolution to left and right, which is not the case in our example. The rough classification produces three types of behavior: periodic behavior,  glider behavior and fractals.
The example we present here would be classified into the third option.

The simplicity of calculations also depends on the initial state being simple to prepare as well. A translation invariant product state as the initial state can be thought of as the result of a quench where we start with the ground state of a ultralocal Hamiltonian for the individual qubits
\begin{equation}
    \hat H = \sum_i \vec v\cdot  \vec \sigma_i 
\end{equation}
for some vector $\vec v$. We then let the system evolve and see how the local degrees of freedom entangle with each other.
The purpose is to understand how the system equilibrates locally for large times in an infinite system and how at finite times one can see similar effects in a finite system with different boundary conditions.

We report that for the class of states described by the initial state $\prod \ket{\psi}$, the system equilibrates for generic $\ket \psi$
in that the 1-qubit density matrix approaches $\rho\to \frac{1}2$ for large times.
There are two behaviors that are important. 
The approach to the limit $\rho\to \frac 12$ is not as uniform as one would imagine. For most times, we get that the 1-qubit density matrix approaches
\begin{equation}
    \rho(t)= \frac 12 +O( \exp(-a t^{h-1}))\label{eq:result1}
\end{equation}
with $a>0$ and dependent on the initial condition that is chosen, and $h= \log_2(3)$, tha Haussdorf dimension of the Sierpinski triangle fractal  (we prove this fact). For times that are powers of $2$, $\rho$ is much farther away from the identity. Namely, the one qubit density matrix near these times only approaches the maximally mixed value polynomially in time
\begin{equation}
    \rho(t=2^k)= \frac 12 +O(\gamma^{k})\label{eq:result2}
\end{equation}
where $\gamma<1$ depends on the initial state. We label this feature a memory effect.

For periodic boundary conditions, we can see this dynamics at play. For open boundary conditions there are additional features.
The system does not thermalize near the boundary, but instead the density matrix has periodic oscillations. We trace these to an additional set of conserved quasi-local operators (they are non-trivial on the left or the right of a point) which only exists in the presence of the boundary. In an infinite line they would be non-local. Here the lack of symmetry on evolution to the left and right relative to the conventional classification of Clifford cellular automata is playing an important role. 
A recent review of quantum cellular automata can be found in \cite{Farrelly:2019zds}.

The paper is organized as follows. In section \ref{sec:CNOTdynamics} we describe the dynamics system in detail. We also explain how the
dynamics on classical bits can be thought of as a linear map.
In \ref{sec:Sierpinski} we understand the evolution of the system in terms of simple classical bit initial states. We note that
the dynamics generated the Sierpinski triangle fractal. We also give a proof of this fact. A similar Sierpinski triangle dynamics arises in the space of operators itself. In section \ref{sec:entanglement}, we study the equilibration dynamics and how it depends on our choice of  
initial states. We give proofs of the results \eqref{eq:result1} and \eqref{eq:result2}. These depend on the fractal dimension of the Sierpinski triangle, and the occasional sparseness of the fractal slice at fixed time for special times. 
We also explain how the (time dependent) conserved
quantities near the boundary are constructed and how they give rise to the oscillation dynamics. We also include simulations for various initial conditions that illustrate these points. The oscillations were first found in the simulations, but it is more clear when explained in the order we choose.

\section{A CNOT gate dynamical system}\label{sec:CNOTdynamics}

The CNOT gate is a simple 2-qubit gate, which acts as follows
\begin{eqnarray}
CNOT \ket{00}& =&\ket{00}\nonumber\\
CNOT \ket{01}& =&\ket{01}\nonumber\\
CNOT \ket{10}& =&\ket{11}\nonumber\\
CNOT \ket{11} &=& \ket{10}
\end{eqnarray}
It takes states in the computational basis of $0,1$ for each qubit, to other states in the computational basis.
This simplicity of the description as a classical computation is what will make quantum  circuits based exclusively on this gate easy to 
evaluate classically.

When we map the states to a column vector as follows
\begin{equation}
\ket{00}\rightarrow\begin{pmatrix}1\\0\\0\\0\end{pmatrix},\quad \ket{01}\rightarrow\begin{pmatrix}0\\1\\0\\0\end{pmatrix}, \quad \ket{10}\rightarrow\begin{pmatrix}0\\0\\1\\0\end{pmatrix},\quad \ket{11}\rightarrow\begin{pmatrix}0\\0\\0\\1\end{pmatrix}\nonumber
\end{equation}
we find that the CNOT gate is characterized by the following unitary matrix
\begin{equation}
CNOT = \begin{pmatrix}
1 & 0 & 0 & 0\\
0 & 1 & 0 & 0\\
0 & 0 & 0 & 1\\
0 & 0 & 1 & 0
\end{pmatrix}    
\end{equation}
There is another way to represent the action of CNOT. The idea is to use the labels of the two qubit kets $\ket{\alpha\beta}$ with $\alpha, \beta \in {0,1}$ to write the set of possible states in a column representation
\begin{equation}
\ket{\alpha\beta} \to \begin{bmatrix}\alpha\\\beta\end{bmatrix}\label{eq:sqbmap}
\end{equation}
where we use the square brackets to distinguish these objects from a column vector.
In this representation, we find that
\begin{equation}
CNOT \begin{bmatrix}\alpha\\\beta\end{bmatrix} = \begin{bmatrix}\alpha\\\beta+\alpha\end{bmatrix} \mod(2).
\end{equation}
This action is then encoded in the matrix representation
\begin{equation}
CNOT\rightarrow \begin{pmatrix}1&0\\
1&1\end{pmatrix}\label{eq:CNOT22}
\end{equation}
The action of CNOT on the basis states \eqref{eq:sqbmap} is equivalent to matrix multiplication by \eqref{eq:CNOT22}, modulo 2.
In this case, we have reduced the original problem of $4\times4$ matrices, to a system of $2\times2$ matrices.

In this paper we will consider quantum computations on a repeated pattern of many qubits, with parallel alternating applications of the CNOT gate between consecutive pairs of qubits, as schematically shown in the circuit \eqref{eq:qcircuit4}:
\begin{equation}
\Qcircuit @C=1em @R=.7em {
& \qw & \ctrl{1} & \qw & \ctrl{1} & \qw & \qw\\
& \qw & \targ & \ctrl{1} & \targ & \ctrl{1} & \qw \\
& \qw & \ctrl{1} & \targ & \ctrl{1} & \targ & \qw\\
& \qw & \targ & \qw & \targ & \qw & \qw
}\label{eq:qcircuit4}
\end{equation}
These systems have translation invariance in time (every two steps of the computation) and they can be thought of as having approximate translation invariance when one is away from the boundaries. 

As such, they can be thought of as modeling a quantum field theory on a lattice with both time and space translation invariance.
If we use a similar notation to \eqref{eq:sqbmap} for $N$ qubits on a line (for simplicity we choose $N$ even), then the first set of parallel gates will 
act as 
\begin{equation}
C1 \begin{bmatrix}
\alpha\\
\beta\\
\gamma\\
\delta \end{bmatrix} =
\begin{pmatrix}
1 & 0 & 0 & 0\\
1 & 1 & 0& 0\\
0 & 0 & 1 & 0\\
0 & 0 & 1 & 1
\end{pmatrix} \begin{bmatrix}
\alpha\\
\beta\\
\gamma\\
\delta \end{bmatrix} \mod(2),
\end{equation}
so we can identify $C1$ with matrix multiplication (modulo $2$)  by a matrix whose entries are either ones or zeros and is given by
\begin{equation}
   C1 \rightarrow  \begin{pmatrix}
1 & 0 & 0 & 0\\
1 & 1 & 0& 0\\
0 & 0 & 1 & 0\\
0 & 0 & 1 & 1
\end{pmatrix} 
\end{equation}
Here, the CNOT gates act on the first and second qubits (indicated by the off-diagonal element in the top $2\times 2$ submatrix), and another gate on the third and fourth qubits. Each CNOT gate is a $2\times 2$ submatrix identical to \eqref{eq:CNOT22}. To have a CNOT in the other direction, 
we would need a sub-block of the type 
\begin{equation}
 {CNOT}_{i+1,i} \rightarrow \begin{pmatrix}
 1&1\\0&1
 \end{pmatrix} 
\end{equation}

The second set of gates will similarly act as 
\begin{equation}
C2\rightarrow
\begin{pmatrix}
1 & 0 & 0 & 0\\
0 & 1 & 0& 0\\
0 & 1 & 1 & 0\\
0 & 0 & 0 & 1
\end{pmatrix} 
\end{equation}
on the computational basis states defined by the square bracket notation objects.
The full circuit dynamics will then be defined by the operator
\begin{equation}
U \equiv C2.C1
\end{equation}
which acts on the basis states modulo $2$. What we will care about is the dynamical system defined by the iteration of $U$ on an initial state.

That is
\begin{equation}
    U^n \ket{\psi_0}
\end{equation}
For $N$ qubits, we then get that 
\begin{equation}
    C1\rightarrow \begin{pmatrix} 
    1&0&0&0&0&\dots\\
    1&1&0&0&0&\dots\\
    0&0&1&0&0&\dots\\
    0&0&1&1&0&\ddots\\
    0&0&0&0&1&\ddots\\
    \vdots &\ddots&\ddots &\ddots &\ddots &\ddots
    \end{pmatrix}
\end{equation}
while
\begin{equation}
    C2\rightarrow \begin{pmatrix} 
    1&0&0&0&0&\dots\\
    0&1&0&0&0&\dots\\
    0&1&1&0&0&\dots\\
    0&0&0&1&0&\ddots\\
    0&0&0&1&1&\ddots\\
    \vdots &\ddots&\ddots &\ddots &\ddots &\ddots
    \end{pmatrix}
\end{equation}
There are additional symmetries. These follow from $C1.C1=1$ and $C2.C2=1$. One therefore gets that $(C12.C1)^-1 = C2.C1 $.
We see this way that $C2.C1= C1. C1. C2.C1= C1. (C2.C1)^{-1}. C1$, so that in this case our dynamics is unitarily equivalent to its time reversal.

We can also have periodic boundary conditions, represented for example in \eqref{eq:qcircuit4p}
\begin{equation}
\Qcircuit @C=1em @R=.7em {
& \qw & \ctrl{1} & \qw & \targ & \ctrl{1} & \qw & \targ&\qw\\
& \qw & \targ & \ctrl{1} & \qw& \targ   & \ctrl{1} & \qw &\qw\\
& \qw & \ctrl{1} & \targ &\qw& \ctrl{1} & \targ & \qw &\qw\\
& \qw & \targ &\qw& \ctrl{-3} & \targ & \qw &\ctrl{-3} &\qw
}\label{eq:qcircuit4p}
\end{equation}
The periodicity then modifies $C2$ slightly by the action of the extra CNOT gate, which adds an extra entry of $1$ in the upper right corner for $C2$
\begin{equation}
C2\rightarrow
\begin{pmatrix}
1 & 0 & 0 & 1\\
0 & 1 & 0& 0\\
0 & 1 & 1 & 0\\
0 & 0 & 0 & 1
\end{pmatrix} 
\end{equation}
The system of Unitaries of $2^N\times 2^N$ matrices acting on states of the $N$ qubit system, has this way been reduced to iteration of a linear system of $N\times N$ matrices modulo $2$. 

\section{The Sierpinski Triangle}\label{sec:Sierpinski}

\begin{figure}[ht]
\includegraphics[width=3.0in]{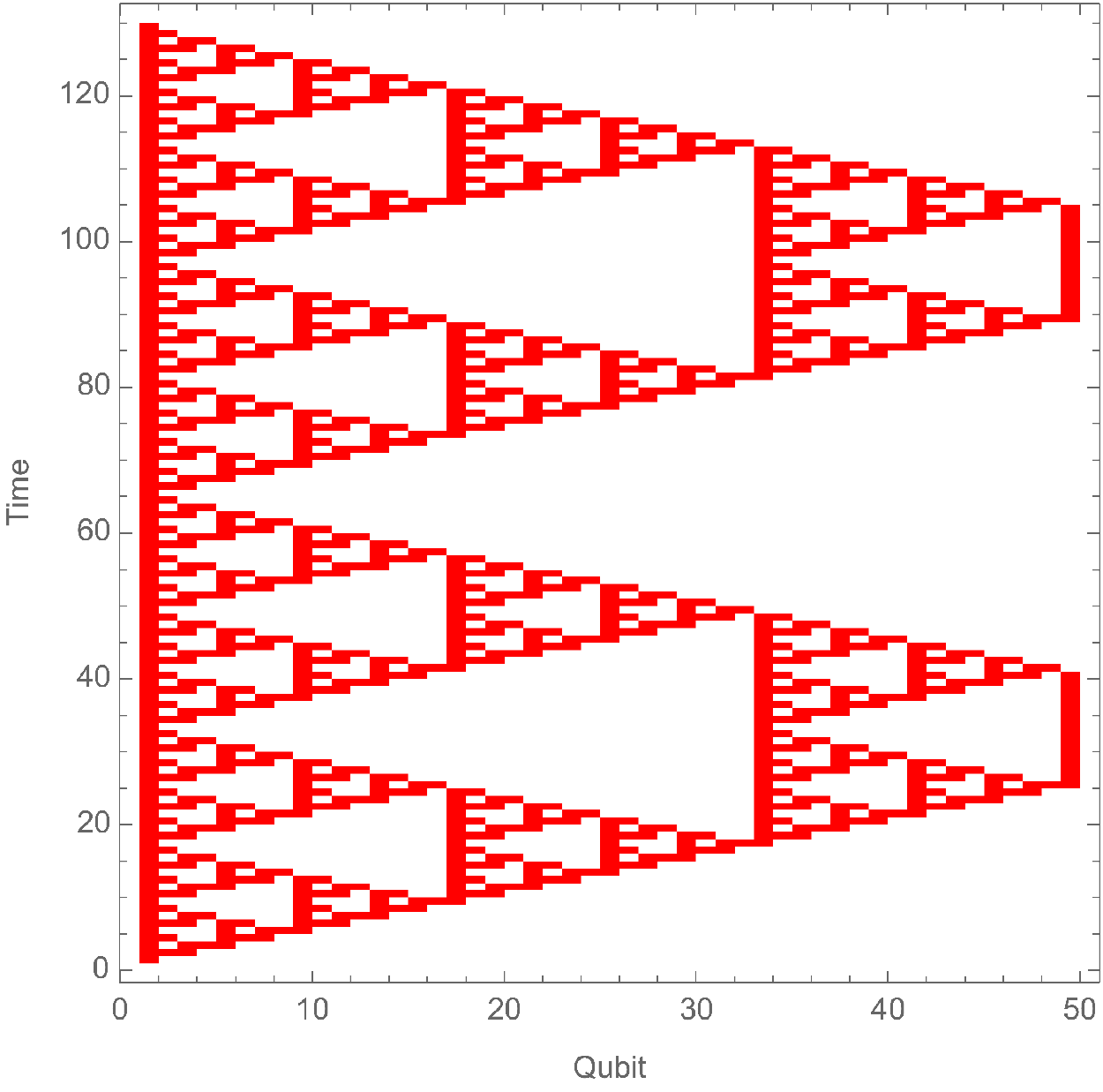}
\caption{Images of $\ket{100}\dots$ under the dynamics. The bits that are turned to the $1$ position are indicated in red.\label{fig:Sierpinski}}
\end{figure}

To begin analyzing the dynamics, let us consider the simplest state with non zero entries, given by 
\begin{equation}
  \ket{\psi_0}=  \ket{10000\dots }\rightarrow \begin{bmatrix}1\\0\\ \vdots \end{bmatrix}
\end{equation}
and let us plot the images $U^n \ket{\psi_0} $ by highlighting the bits that get turned on at time $t$. We get the  picture in figure \ref{fig:Sierpinski}. We see a figure that looks like the Sierpinski triangle. Notice the symmetry of the figure about time $t=64$. This is a
reflection of the time reversal symmetry of the dynamics.

For other starting bits, the pattern is similar, shifted to the right.
We will now give a derivation of the fact that we get the Sierpinski triangle in the results above.
The idea is to shift the order in which the gates are applied.
Consider instead the following order
\begin{equation}
 \Qcircuit @C=1em @R=.7em {
& \qw & \qw & \qw &\ctrl{1}&\qw & \qw & \qw &\ctrl{1}& \qw\\
& \qw  & \qw & \ctrl{1}  &  \targ  & \qw & \qw & \ctrl{1}  &  \targ &\qw \\
& \qw   & \ctrl{1} & \targ & \qw  & \qw  & \ctrl{1} & \targ & \qw & \qw\\
& \qw & \targ & \qw  & \qw & \qw & \targ & \qw  & \qw & \qw
}
\end{equation}
and  repeat this circuit instead of with the new unitary $\tilde U$ instead of the one in \eqref{eq:qcircuit4}.

It is easy to see that the state $\ket{100\dots}$ will propagate essentially the same way in the new circuit than in the old one, except that the
time and space have been sheared relative to each other. We are allowed to do this shift in the order of the operations because the operators that we are shifting  commute with each other. 
After all, they act on different qubits \footnote{Note that this shift can not be carried out with 
periodic boundary conditions: the operators eventually wind around the circle.}.
 This different time slicing breaks the time reversal symmetry.
Apart from the one qubit that is initialized in the $\ket 1$ state, all the other that start at zero are stable (stay zero) when we do the shearing operation.

This corresponds to a new matrix given by
\begin{equation}
\tilde U 
\rightarrow 
\begin{pmatrix} 1& 0 & 0 & \dots\\
1&1&0&\dots\\
0&1&1& \ddots\\
\vdots& \ddots & \ddots & \ddots
\end{pmatrix}
\end{equation}
In this dynamical system we have that at time $n$ the kets are given by elements of the computational basis of type
$\ket{1 \alpha_{1n} \alpha_{2,n} \dots}$, with 
\begin{eqnarray}
\alpha_{0,n}&=&1 \nonumber \\
\alpha_{i,0}&=& 0  \hbox{ for } i\neq 0, \nonumber \\
\alpha_{i,n}&=& (\alpha_{i,n-1}+\alpha_{i-1,n-1}) \mod (2),  \quad \hbox{ for } i\neq 0
\end{eqnarray}
The first two detail initial conditions. The last one gives a function that updates the $n$-th qubit depending on the state at the previous time in its position, and one of its neighbors. This is one of the elementary  cellular automata (rule 60), which is known to give rise to the Sierpinski triangle fractal pattern. Unlike the infinite cellular automata, where the state $\dots 00 0\dots $ can be arrived at from both the state  $\dots 00 0\dots $ and $\dots 111 \dots $,
the evolution of the quantum circuit takes a single state to a single state. This memory loss would only appear in an infinite volume limit in both directions. The slicing argument would not automaticallly lead to an equivalence because there is an infinite product of operators in the reshuffling. These can be pathological.

\begin{figure}[ht]
\includegraphics[width=3.0in]{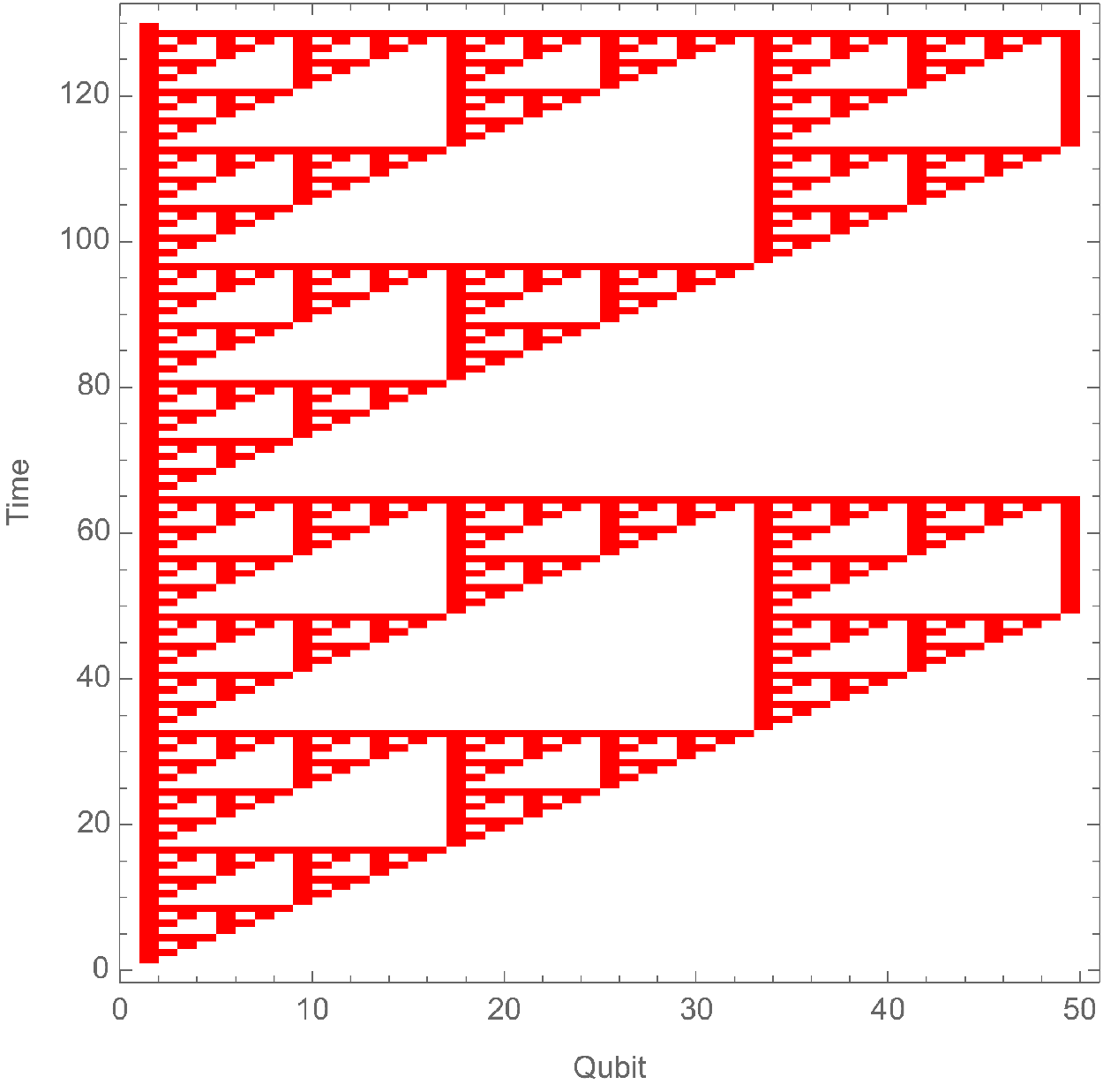}
\caption{Images of $\ket{100\dots}$ under the $\tilde U$ dynamics. This is rule 60 dynamics for a cellular automaton. The bits that are turned to the $1$ position are indicated in red \label{fig:Sierpinskishear}.}
\end{figure}

One can easily check that the recursion relation is solved by
\begin{equation}
\alpha_{n,i} = {n \choose i} \mod(2)    
\end{equation}
which is one of the  definitions of the Sierpinski triangle: the binomial coefficients modulo 2.
 The graph we get in this case is given by figure \ref{fig:Sierpinskishear}, which is clearly a sheared version of the figure
 \ref{fig:Sierpinski}.

If we start with a classical configuration (a fixed set of values in the computational basis), then the evolution of the system will follow the classical evolution of the dynamics. 
The only new ingredient brought about by quantum mechanics is the possibility of studying non-trivial superpositions of states.
Our ultimate goal is to understand the dynamics of the quantum version of the system for different classes of quantum initial conditions.

\subsection{Sierpinski triangle in Operator dynamics}

Just like in the evolution  of states we see the Sierpinski triangle, it turns out hat it also shows up in the operator dynamics.
When we writethe dynamics in the Heisnberg representation, the states are fixed and the operators change,. For each qubit we have 
a list of four standard operators
\begin{equation}
    1, X_j, Y_j, Z_j \simeq 1, \sigma_x, \sigma_y, \sigma_z
\end{equation}
The Heisenberg evolution will be of the form
\begin{equation}
    Z_j(t) = U^{-t} Z_j(0) U^{t}
\end{equation}
and similar for $X,Y$.

It is easy to check that (since $CNOT^2=1$) that the following is true
\begin{eqnarray}
    (CNOT)_{j-1,j}  Z_{j-1}  (CNOT)_{j-1,j} &=& Z_{j-1}\\
     (CNOT)_{j-1,i}  Z_{j}  (CNOT)_{j-1,j} &=& Z_{j-1}\otimes Z_j
\end{eqnarray}
The dynamics has a subalgebra generated by the $Z$ that is mapped to itself. 
Such Clifford cellular automata with period one (that are not trivial) are not allowed in the classification with periodicity of one step \cite{schlingemann2008structure}. They inevitably mix $X,Z$ and moreover generate symmetric evolution to the left and right \footnote{This can be traced to the classification being built of  palyndromic polynomials with coefficients in ${\mathbb Z}_2$}.

Now, consider the following. To a series of bits $c_1\dots c_N$, we can associate the operator
$Z_1^{c_1}\dots Z_N^{c_N}$, where by convention $Z^0=1$. Due to the action above, the labels $c$ evolve with the same dynamics as
that of the states themselves. This is natural:  to change the value of $Z_j$ at time $t$ we can equally 
change the value of $Z$ at time zero for any of the bits that land on the $j$ bit at time $t$.
The dynamics of the $Z$ is thus the same type of Sierpinski triangle dynamics as we have for the bits and can be computed using linear algebra (modulo 2) on the $c$ sequences.

Similarly, 
\begin{eqnarray}
    (CNOT)_{j-1,j}  X_{j-1}  (CNOT)_{j-1,j} &=&  X_{j-1} \otimes X_j \\
    (CNOT)_{j-1,j}  X_{j}  (CNOT)_{j-1,j} &=&  X_{j
} 
\end{eqnarray}
and to a series of bits $b_1\dots b_N$, we associate the operator
$X_1^{b_1} \dots X_N^{b_N}$ acting on $N$ qubits, where $X^0=1$. We see then that the Heisenberg dynamics induced on these bit sequences by the dynamics
associated to $U$ gives a dynanics that flows in the opposite direction.
As such, $X_1$ will similarly evolve into a copy of the Sierpinksi triangle, but running to the left, rather than the right (relative to figure \ref{fig:Sierpinski}).

This opposite dynamics reflects the fact that the $CNOT_{i-1,i}$ is unitarily equivalent to the $CNOT_{i,i-1}$ gate. These are related to each other by applying 
Hadamard gates on all the qubits before and after the action of the $CNOT$ operator.

From these, one can deduce the action on the $Y \simeq I X Z$ by multiplication of the results for $Z,X$. 
Here we denote $I^2=-1$ the pure imaginary number, so as not to confuse it with possible indices for the $X,Z$ variables.
This set of monomials in the Pauli matrices (up to additional factors of $\pm 1, \pm I $) give rise to the Pauli group. 
Indeed, this dynamics is an example of an operator in the Clifford group, which is the normalizer of the Pauli group.

Indeed, this is an example of the types of systems considered in \cite{Gottesman:1998hu}, whose Heisenberg dynamics is particularly simple and classically computable.

\section{Entanglement entropy dynamics}\label{sec:entanglement}

We can now use the results of the previous section to understand the thermalization dynamics of the quantum circuit, when we start 
the dynamics on the translation invariant product state \footnote{Because the gate dynamics alternate the pairs of qubits, the dynamics is only spatially translation invariant  with period $2$.
}
\begin{equation}
    \ket{\psi}_{in}= \prod \ket {\psi_0}. 
\end{equation}

For a single site, we can compute the reduced density matrix at the site as follows
\begin{equation}
    \rho_i(t) = \frac 12 \left( 1+  X_i \vev{X_i}_t+ Y_i \vev{Y_i}_t+Z_i \vev{Z_i}_t \right)
\end{equation}
where we evaluate the expectation values at time $t$. This is a $2\times 2$ density matrix and from it we can compute the entanglement entropy of the corresponding qubit. The idea for simplifying the computation is to use the Heisenberg representation of the evolution. In this case
the expectation values are of the form
\begin{equation}
\vev{Z_i}_t= \bra{\psi}_{in} Z_i(t) \ket{\psi}_{in} 
\end{equation}
Notice now that 
\begin{equation}
    Z_i(t) = U^{-t} Z_i(0) U^t \simeq \bigotimes Z^{c_{j,i}}_j
\end{equation}
where $b_{i,j}$ is a bit sequence that is the image of $Z_i$ as a product of $Z_j$ and where $Z^0=1$. 
Because the initial state is a product state, we can then evaluate directly 
\begin{equation}
    \vev{Z_i(t)} = \prod_j \bra{\psi_0} Z \ket{\psi_0}^{c_{i,j}} = \bra{\psi_0} Z \ket{\psi_0}^{\sum c_{i,j}}\label{eq:prodform}
\end{equation}
and the right hand side is readily computable.
Similarly
\begin{equation}
    \vev{X_i(t)} = \prod_j \bra{\psi_0} X \ket{\psi_0}^{b_{i,j}} = \bra{\psi_0} X \ket{\psi_0}^{\sum b_{i,j}}
\end{equation}
The numbers $c_{i,j}$ and $b_{i,j}$ are readily computable.

Finally, using $Y_i= I X_i Z_i $, we get that 
\begin{equation}
    \vev{Y_i(t)} =  \bra{\psi_0} Y \ket{\psi_0} \prod_{j \neq i} \bra{\psi_0} X \ket{\psi_0}^{b_{i,j}} \bra{\psi_0} Z \ket{\psi_0}^{ c_{i,j}}
\end{equation}
where (for open boundary conditions) we are using the property of the dynamics that the $X_i(0)$ can only affect $X_{j\leq i}(t)$, while $Z_i(0)$ can only affect $Z_{j\geq i}(t)$.
In the case of periodic boundary conditions we need to be a bit more judicious, as the $X,Z$ images can readily interfere with each other after wrapping the circle. Every time $c_{j,i}= b_{j,i}=1 $, we get an expectation value  $-I\vev Y$ in the product. 
From here, it is straightforward to compute the entropy of the corresponding qubit.

Generalizing this computation to more qubits seems straighforward: we use a basis for operators made of products of $1,X,Y,Z$ at each site and 
combine them. Similar sequence of bits will arise. It is still expected that the operators would superpose the Sierpinski triangles in some form, qualitatively giving results like in the right hand side.
The harder part would be to compute it in practice, as the various $X,Y,Z$ interfere with each other.

We will now show that the system thermalizes at large times  to a maximally mixed density matrix (for single qubits) for most initial states of the classes we have considered, when we take the thermodynamic limit. At finite size, the dynamics is periodic, either with periodic boundary conditions or with open boundary conditions. 

\begin{figure}[ht]
\includegraphics[width=3.0in]{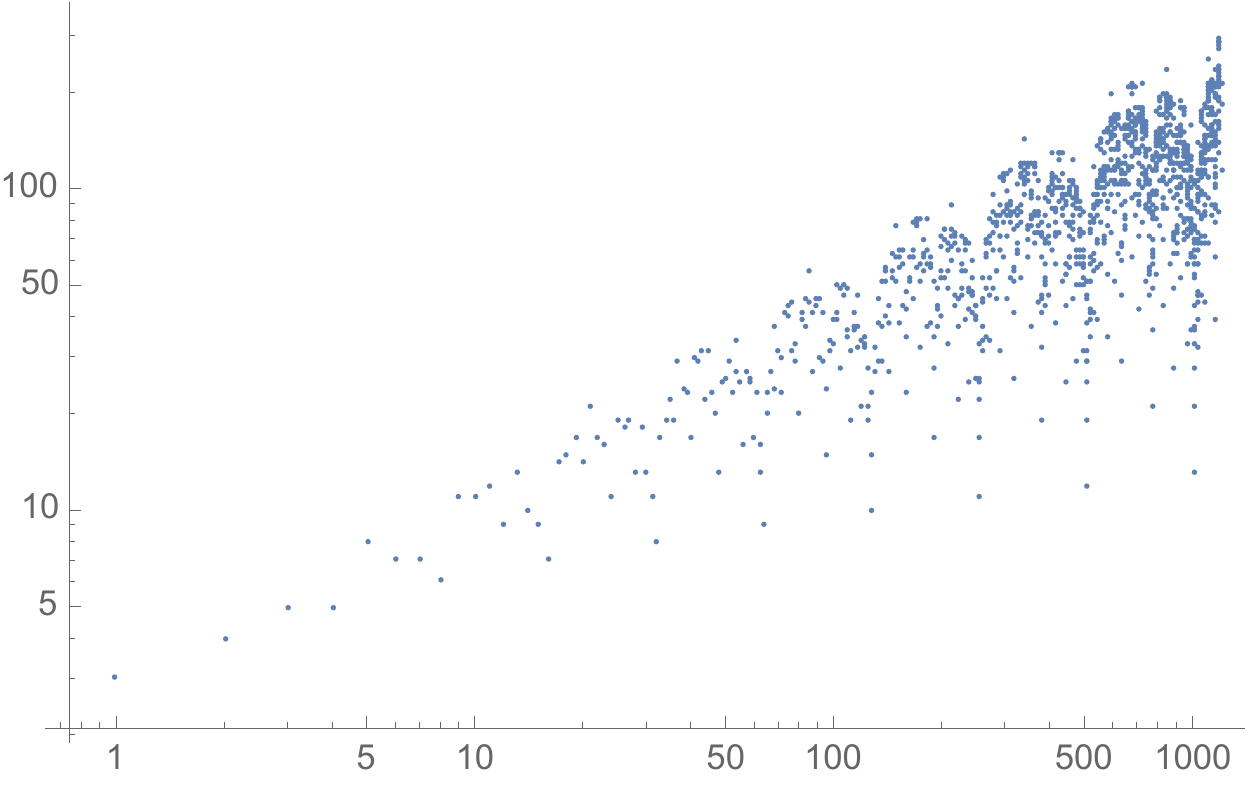}
\caption{Number of bits turned on in the images of $10\dots0$ as a function of time, in the dynamics depicted in figure {\ref{fig:Sierpinski}}\label{fig:Sierploglog}}
\end{figure}

The basic idea is that $\vev{X_i(t)}\to 0$ as $t\to \infty$, and similarly for $Y$, $Z$. This follows so long as $x=|\vev X|<1$, $z=|\vev{Z}|<1$.
As we can see, in the right hand side of \eqref{eq:prodform}, we get $(\vev{X})^{\sum b_{ij}(t) }$. We can easily see that the sums appearing in the exponent grow 
as $t\to\infty$. This is also true for $Z$. This is illustrated in figure \ref{fig:Sierploglog}.

The figure is presented in a log-log plot. The reason for doing the plot this way is that we clearly see that the number of bits turned on follows a scaling relation. We also see some periodic features (dips in the number of bits turned on) in $\log(t)$. Because the Sierpinski triangle is fractal, there is a non-trivial Haussdorf dimension at play. 
The Haussdorf dimension of the triangle is the logarithm of number of points in the triangle at time $t$, relative to the size of the triangle. That is $h\simeq \lim_{t\to \infty}\log(\# Points)/Log(t)$. Thus, roughly, the cumulative number of bits turned on
at time $t$ scales like $t^h$.
The number of bits that are on exactly at time $t$ scales on average like $t^{h-1}$ where $h$ is the Hausdorff dimension of the triangle. That way, the area of the bits that are turned on in the full triangle scales like $t^{h} \simeq \int t^{h-1} dt$. In this case, the Haussdorf dimension $h$ is known and given by  $h= \log(3)/\log(2)= 1.584\dots$.
This means that the density matrix approaches the identity with a non trivial time dependence
\begin{equation}
 \rho \to \frac 12 +O( \exp(-s t^{h-1}))
\end{equation}
Notice that this thermalization is slower than exponential decay and that this is controlled by the Hausdorff dimension of the fractal.
The dependence on the state is measured by the possible $s$ values. Generically, the numbers  $x,z \simeq \exp(-s)$ encode $s$, 
or $xz \simeq \exp(-s)$ for $Y$ equilibration. 

On the other hand, the log-periodic dips indicate that the system has memory flashbacks where the entropy can be very low (the system is less close to equilibrium at the corresponding qubit). This is especially noticeable when either $x$, or $z$ is very close to $1$. In that case the system starts with a state that is very close to the computational basis (or its discrete Fourier transform counterparts).

Indeed, these correspond exactly to times that are  powers of $2$, namely $t=2^m$. These are the vertices where the largest empty triangle tips  meet the vertical axis in figure \eqref{fig:Sierpinski}. The number of bits turned on in that case is only $m+3$. For these special times (which asymptotically are of measure zero), we have that 
\begin{equation}
 \rho \to \frac 12 +O( \exp[-\alpha\log(t)])
\end{equation}
so that the approach to equilibrium at flashback times is only power law. The power law depends on the initial state.

For the special case where $\vev X= \vev Z=0$, that is, an eigenstate in the $Y$ basis,  we see that $\vev{Y} =0$ in the future for all times. Equilibration to the maximally mixed state (density matrix) for the qubit is instantaneous. 

When we have open boundaries, the analysis changes near the boundaries. That is because the sequences that run to the left terminate, in that no new qubits can be found to the left  (similar to the right). So for example $Z_1$ is conserved, as well as its expectation value. In that sense, the density matrix does not converge to the maximally mixed density matrix. Instead, only the off-diagonal elements converge to zero.
Indeed, the subalgebras generated by $\{Z_1,  \dots,  Z_k\}$ are all preserved for each $k$. 
This is a type of generalized conservation law that forbids thermalization near the edges. Instead, we expect that the local density matrix will depend on the expectation values of the elements of the subalgebra. 
We can actually find conserved charges for this fact. For example,  $Z_2$ will evolve to $Z_1\otimes Z_2$ and back. Thus the average of the two is conserved, and their difference oscillates in sign with period 2, Namely
\begin{equation}
s_2=  \left\langle   \frac 12(Z_2+Z_1\otimes Z_2)\right \rangle\quad \tilde s_2=  (-1)^t\left\langle   \frac 12(Z_2-Z_1\otimes Z_2)\right \rangle
\end{equation}
are both time independent. That means that $\vev {Z_2}= s_2 +(-1)^t \tilde s_2$ is oscillatory. The entanglement entropy itself must oscillate as
well with these periods even after a notion of equilibration has taken hold. A similar construction will follow for $Z_3$ etc. Let us be explicit. Let us call the map of the $Z$ into each other by $\mu$. Then
\begin{eqnarray}
\mu(Z_1) &=& Z_1\\
\mu(Z_2) &=& Z_1\otimes Z_2\\
\mu(Z_3) &=& Z_1\otimes Z_2\otimes Z_3\\
\mu(Z_4)&=& Z_3\otimes Z_4\\
\mu(Z_5)&=& Z_3\otimes Z_4\otimes Z_5
\end{eqnarray}
We see generally that the map $\mu(Z_k)$ only contains the $Z_s$ for $s\leq k$. 
We can average over the orbit of $Z_3$ to find a conserved charge. We can also do a discrete Fourier transform on the images of $Z_3$ with respect to time to find the oscillating pieces. From here, we can reconstruct the expectation value of $Z_3$ and similarly for others.
Let us compute directly
\begin{equation}
\mu(\mu(Z_3)) = \mu(Z_1) \mu( Z_2) \mu(Z_3) = Z_1^3\otimes Z_2^2\otimes Z_3=Z_1\otimes Z_3
\end{equation}
It is easy to check that $\mu^4(Z_3)= Z_3$. Consider a root of unity $\omega$, such that $\omega^4=1$. The four (oscillating) conserved quantities are then given by
\begin{equation}
I_3(k ) = \sum_{i=0}^3 \omega^{-i *k}  \mu^i(Z_3)
\end{equation}
and it is easy to show that $\mu(I_3(k))= \omega^k I_3(k)$.
The expectation values of $I_3(k) \omega^{-k t}$ are time independent. We can reverse the discrete Fourier transform to find that then the expectation value of 
any product of the $Z_i$ for $i\leq 3$ is also periodic.   

In the thermodynamic limit for the boundary (where the other end is very far away),
we expect local oscillations on the one qubit entropy for each qubit near the boundary, as well as any element of the density matrix of the $k$ qubits near the end, that only depends of expectation values of products of $Z_i$, for $i\leq k$.

The operators $I_3(k)$ (and their generalizations to arbitrary $k$) play the role of the $A$ operators in \cite{s41467-019-09757-y} in the continuous time framework: local operators whose time evolution is as an eigenstate of the Liouvilian operator.  Density matrices constructed from these are time periodic and correspond to the long time limit behavior of the system. Systems that have such ``non-local" conservation laws generally exhibit oscillations \cite{PhysRevB.102.041117}. Such oscillations are persistent also from the point of view of OTOC's (\cite{PhysRevX.8.021013}) in finite volume. Here the effect exists even at infinite volume with a boundary.

For the $X$ variables, it is $X_N$ that is conserved, and the right nested subalgebras generated by  $\{X_{N-k} \dots X_N\}$. 
Both of these indicate that the boundaries should be equilibrating to a more generalized ensemble adapted to the nested algebras. 

\subsection{Open boundary conditions}

\begin{figure}[ht]
    \centering
    \includegraphics[width=2.9in]{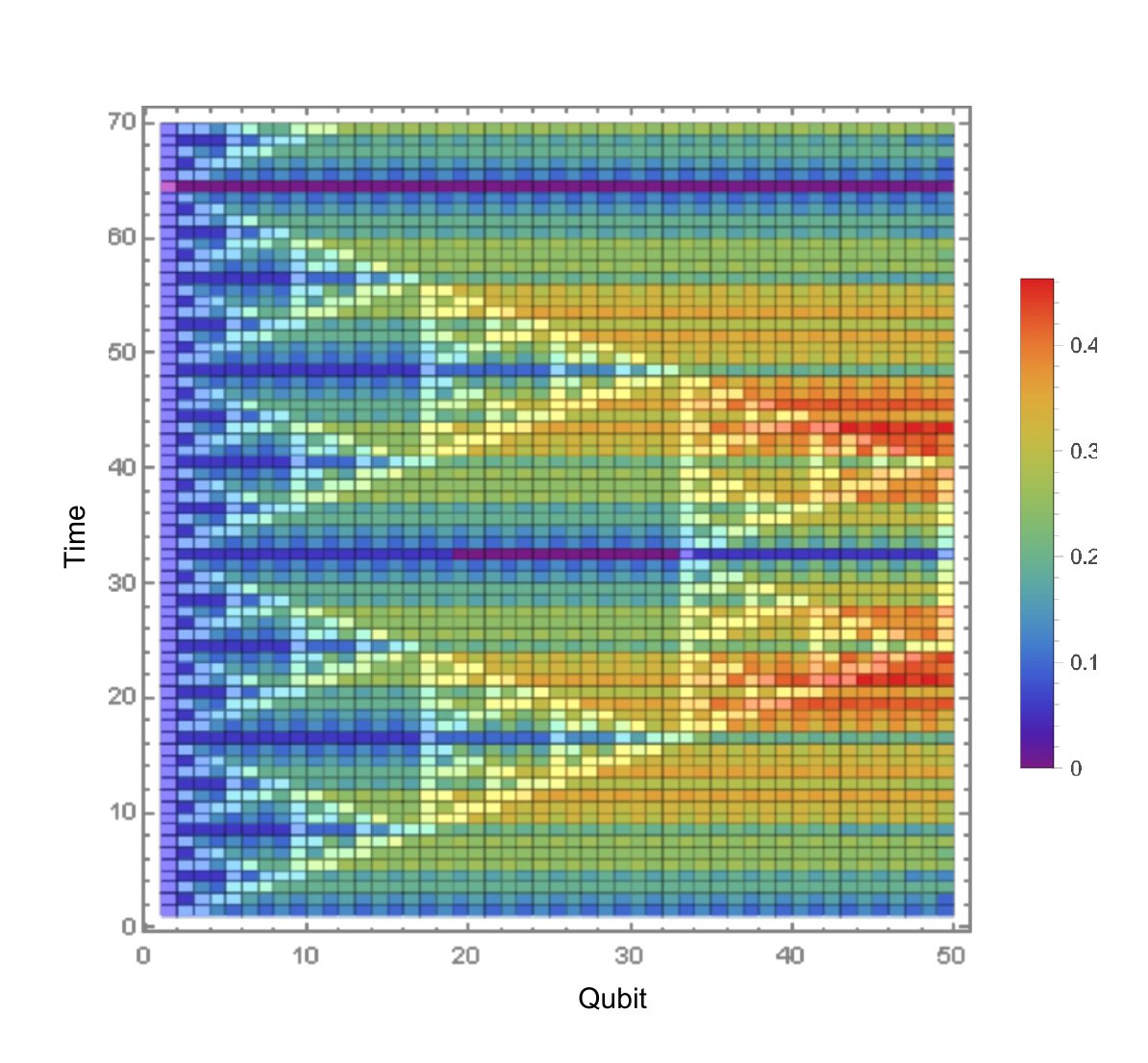} \includegraphics[width=3.0in]{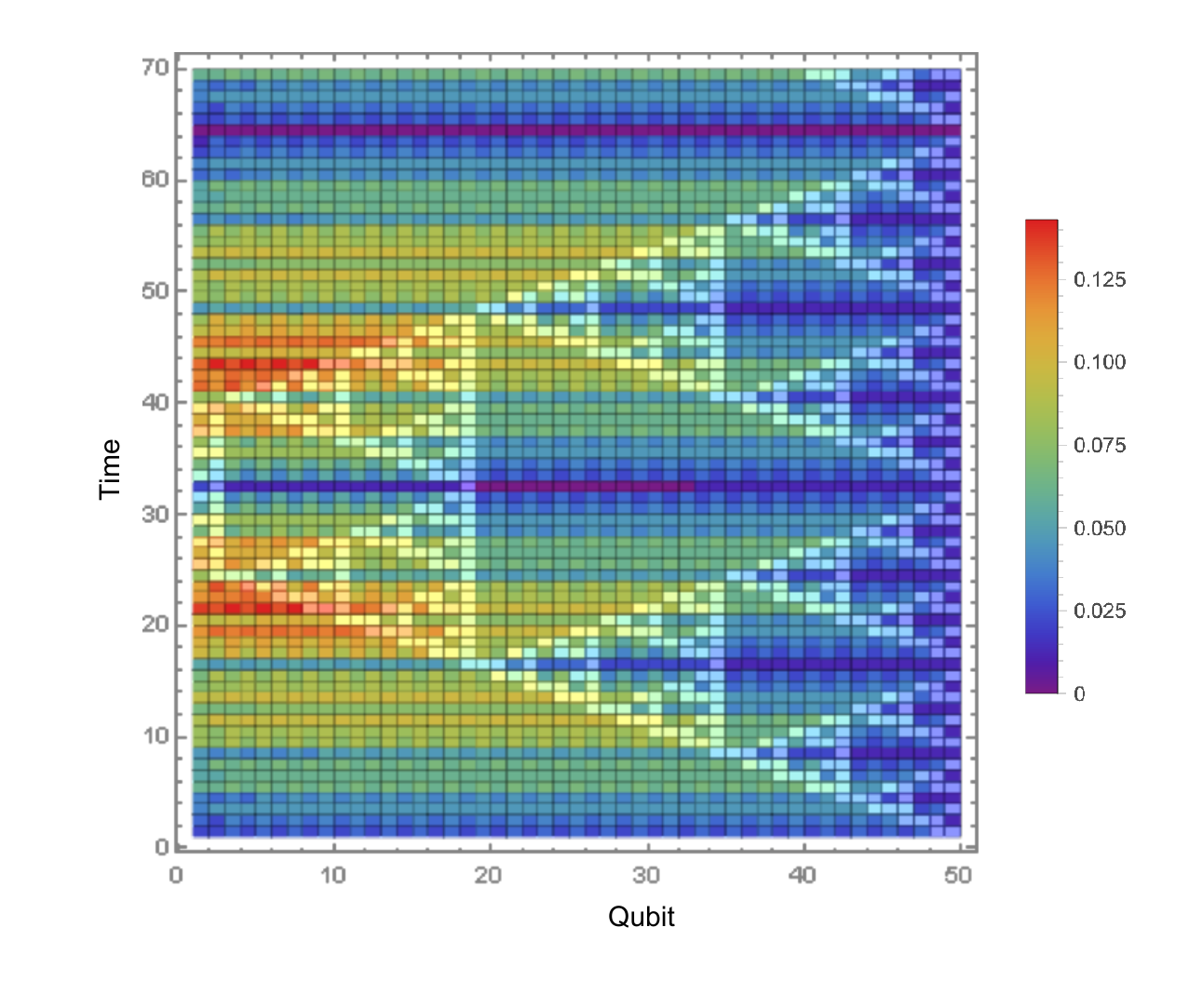}
    \caption{Single qubit entanglement enrtropy for two different initial conditions on a system of 50 qubits. On the left, we start with $\ket{\psi_0}\sim  0.99 \ket 0+ 0.1 \ket 1$, close to the $Z$ eigenstate. In
    the right, we have $\ket{\psi_0}\sim 0.735 \ket 0+ 0.678 \ket 1$. The Sierpinski triangle of the leftmost $X$ is superimposed on the left, 
    while the Sierpinski triangle  of the rightmost $Z$ is superimposed on the right. This is done to graphically show that the entanglement entropy tracks the triangle.} 
    \label{fig:ent1q_open}
\end{figure}

Let us now show some examples of the entanglement entropy for single qubits in the case of open boundary conditions.
This is depicted in figure \ref{fig:ent1q_open}.
As can be clearly seen, in the left figure the qubits on the left have relatively low entropy, and the entropy increases towards the right. 
The largest value of the entropy occurs when the number of bits in the Sierpinski triangle is large. A zone with zero entropy is seen in the center of the figure at time $t=16$, where the number of points of either Sierpinski triangle misses the center.
These are qubits whose images
under the evolution map get mapped outside the interval. This is similarly seen on the right hand side. 
We notice that the figures are essentially mirror images of each other.
For each of these, the time evolution is periodic with period $t=64$, and the system returns to zero entropy for all qubits at that time.
Similarly, we see dips in the entropy near the boundary at other times $t=16$ and at even times (depending on the largest power of 2 that divides the time). These features seem approximately periodic depending on the size of the subregion, as expected from our theoretical observations at the beginning of the section. For the state near the $Z$ axis, we see that the entropy on the left qubits is small, as predicted by our arguments at the beginning of the section: these are due to the state being close to an eigenstate of the left local subalgebras of the $Z$.  We see similar features for the qubits on the right when we are close to an $X$ eigenstate..

\begin{figure}[ht]
    \centering
    \includegraphics[width=3.4 in]{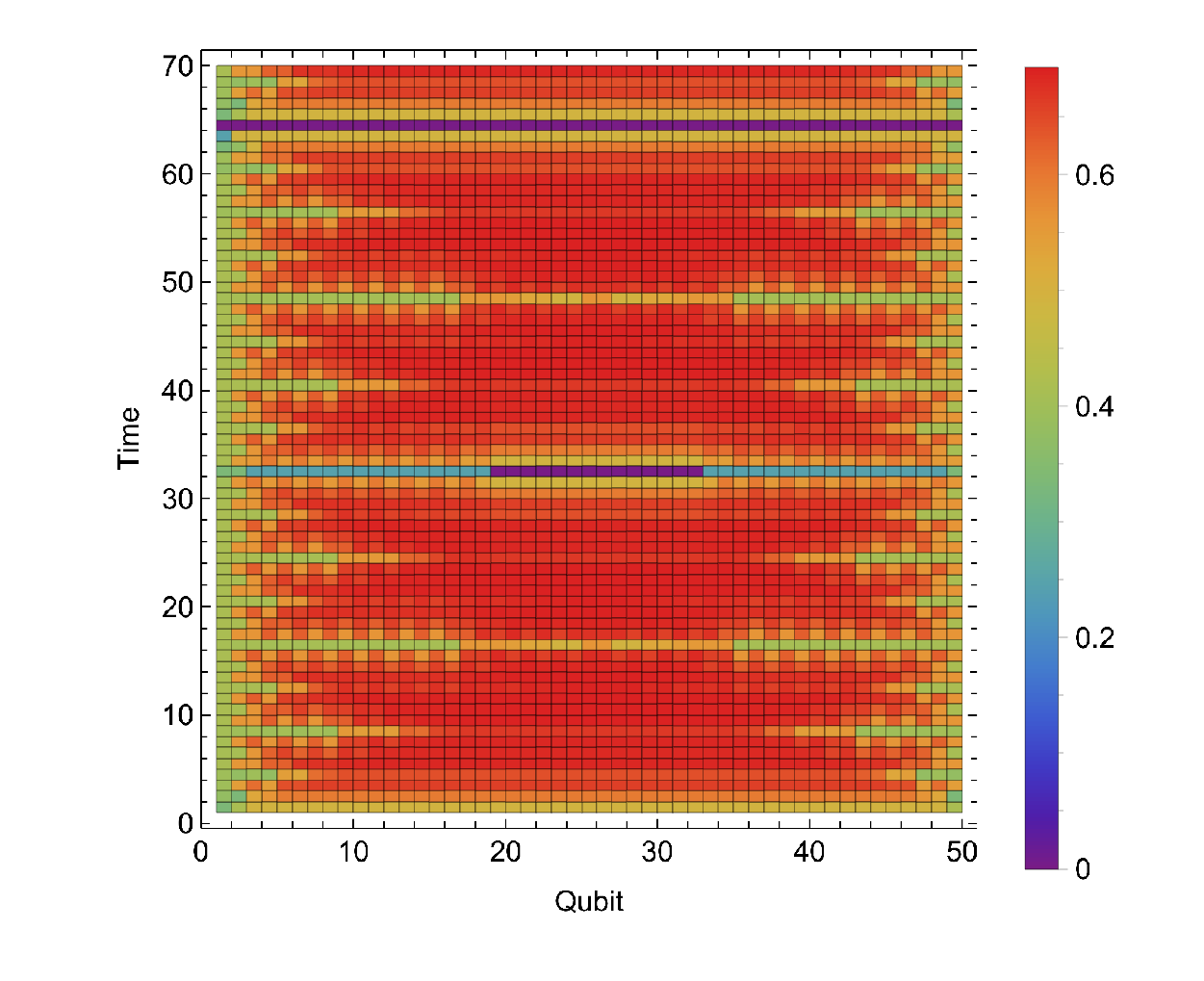}\includegraphics[width=3.0 in]{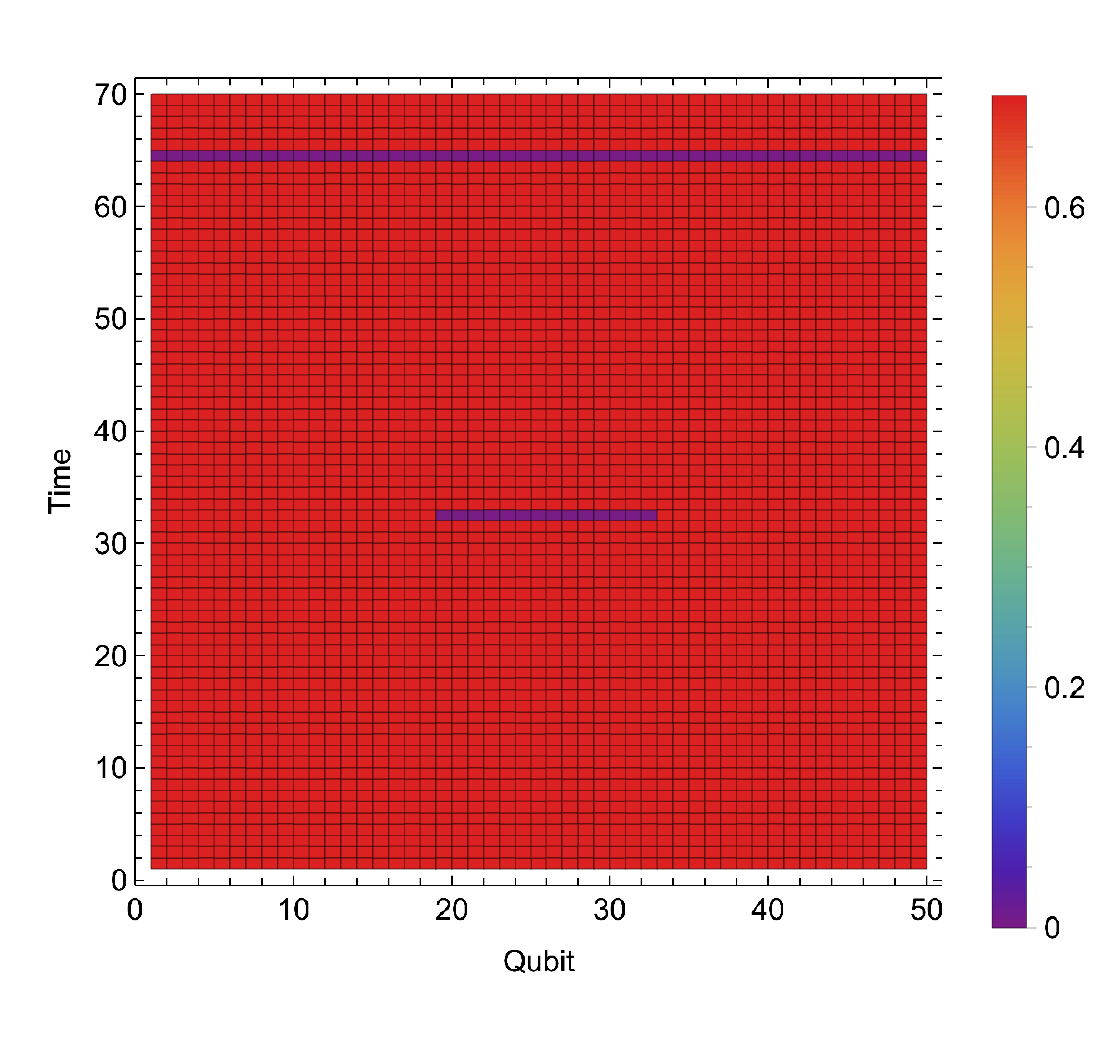}
    \caption[width=4 in]{Entanglement entropy of single qubits: symmetric real point between X and Z basis on the left. Y eigenstate on the right.}
    \label{fig:50_symmetric}
\end{figure}

We can now also go to the halfway point between the $Z$ and $X$ basis states with real coefficients. This is depicted in \ref{fig:50_symmetric}.
The figure shows the symmetry on reflection about the middle. The center also has a region with zero entropy, just as before, 
We also have Y eigenstate initial state. This shows that the entropy goes maximal in one shot for all qubits. There is still the slit on the center at the half time of the period.

Just like we can write a density matrix for a single qubit, we can do so for many qubits to understand the entanglement paterns that appear in the dynamics. For two qubits, this is given by
\begin{eqnarray}
    \rho&=& \frac 14 ( 1 + 1\otimes X \vev{1\otimes X} +   1\otimes Y \vev{1\otimes Y}+1\otimes Z \vev{1\otimes Z}\nonumber\\
    &+& X\otimes 1 \vev{X\otimes 1} + X\otimes X \vev{X\otimes X} +   X\otimes Y \vev{X\otimes Y}+X\otimes Z \vev{X\otimes Z}\nonumber\\
     &+& Y\otimes 1 \vev{Y\otimes 1} + Y\otimes X \vev{Y\otimes X} +   Y\otimes Y \vev{Y\otimes Y}+Y\otimes Z \vev{Y\otimes Z}\nonumber\\
     &+& Z\otimes 1 \vev{Z\otimes 1} + Z\otimes X \vev{Z\otimes X} +   Z\otimes Y \vev{Z\otimes Y}+Z\otimes Z \vev{Z\otimes Z}
\end{eqnarray}
and we can evaluate the expectation values at any time. Just like for a single qubit, the expectation values can be written as various products
of elementary expectation values $\vev {X(0)}, \vev{ Z(0)}, \vev{ Y(0)}$.

\begin{figure}[ht]
    \centering
    \includegraphics[width=3.3in]{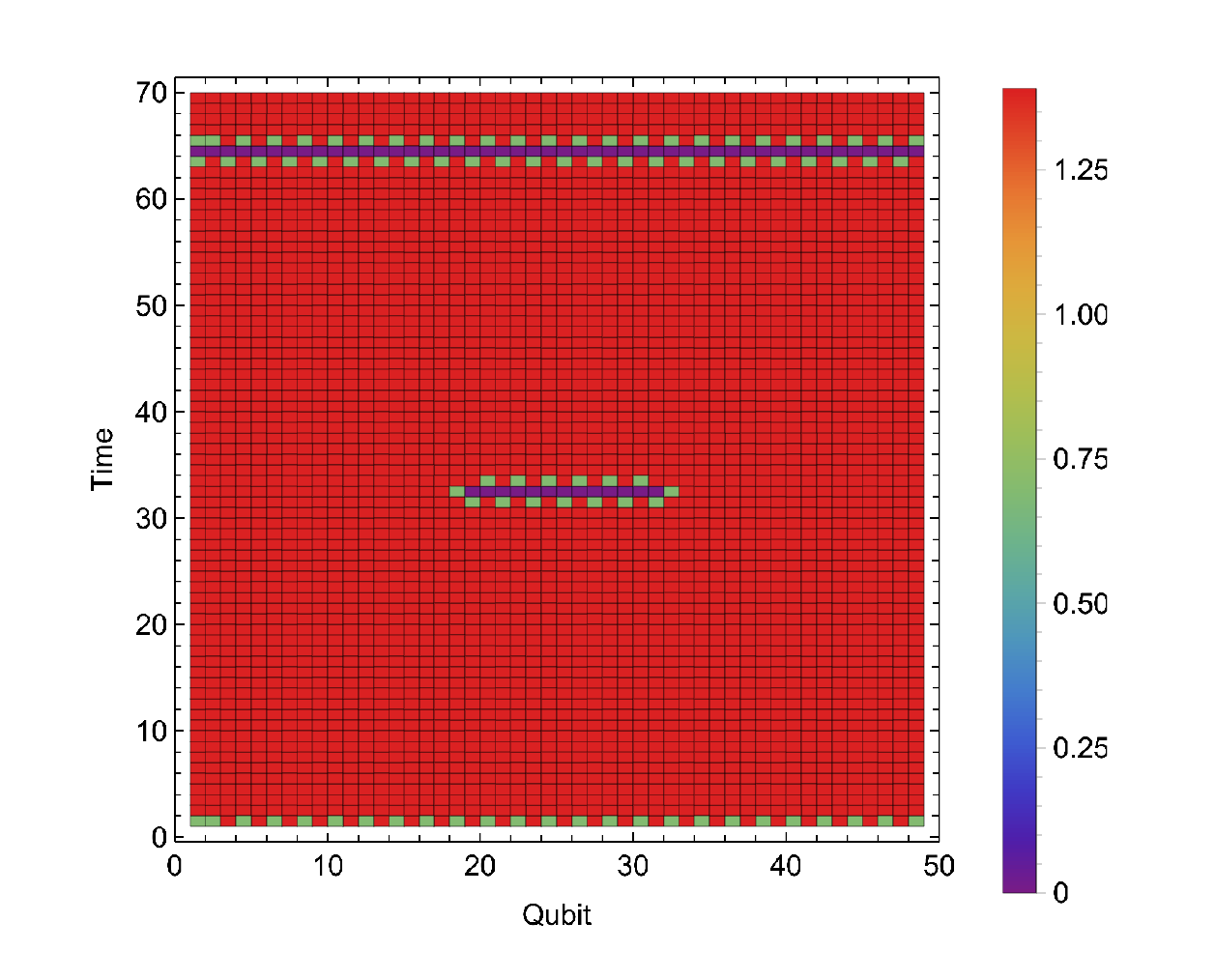}\includegraphics[width=3in]{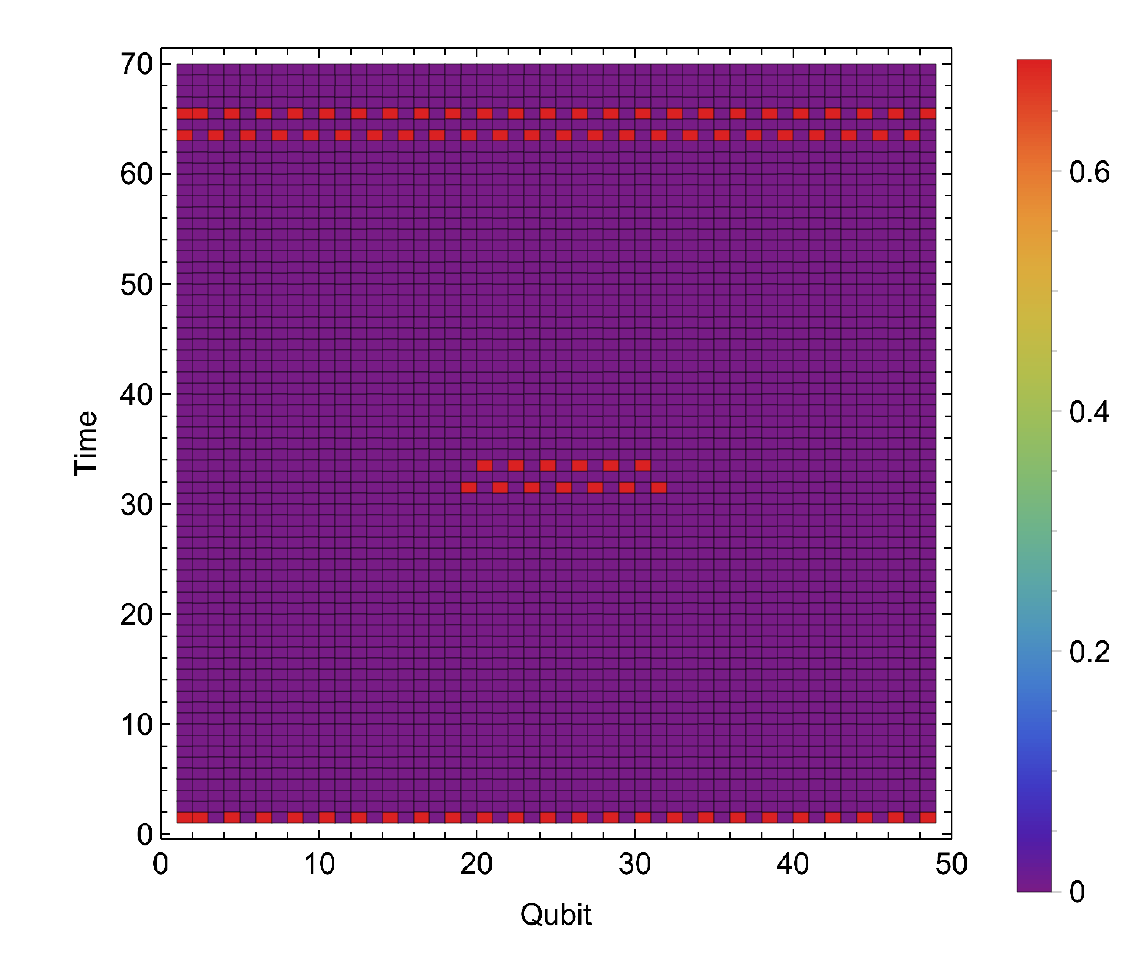}
    \caption{Total entropy for pairs of nearest neighbors for Y eigenstate initial conditions on the left. Mutual information for nearest neighbors in the same initial conditions on the right. }
    \label{fig:MI_openY}
\end{figure}

\begin{figure}[ht]
    \centering
    \includegraphics[width=4.5in]{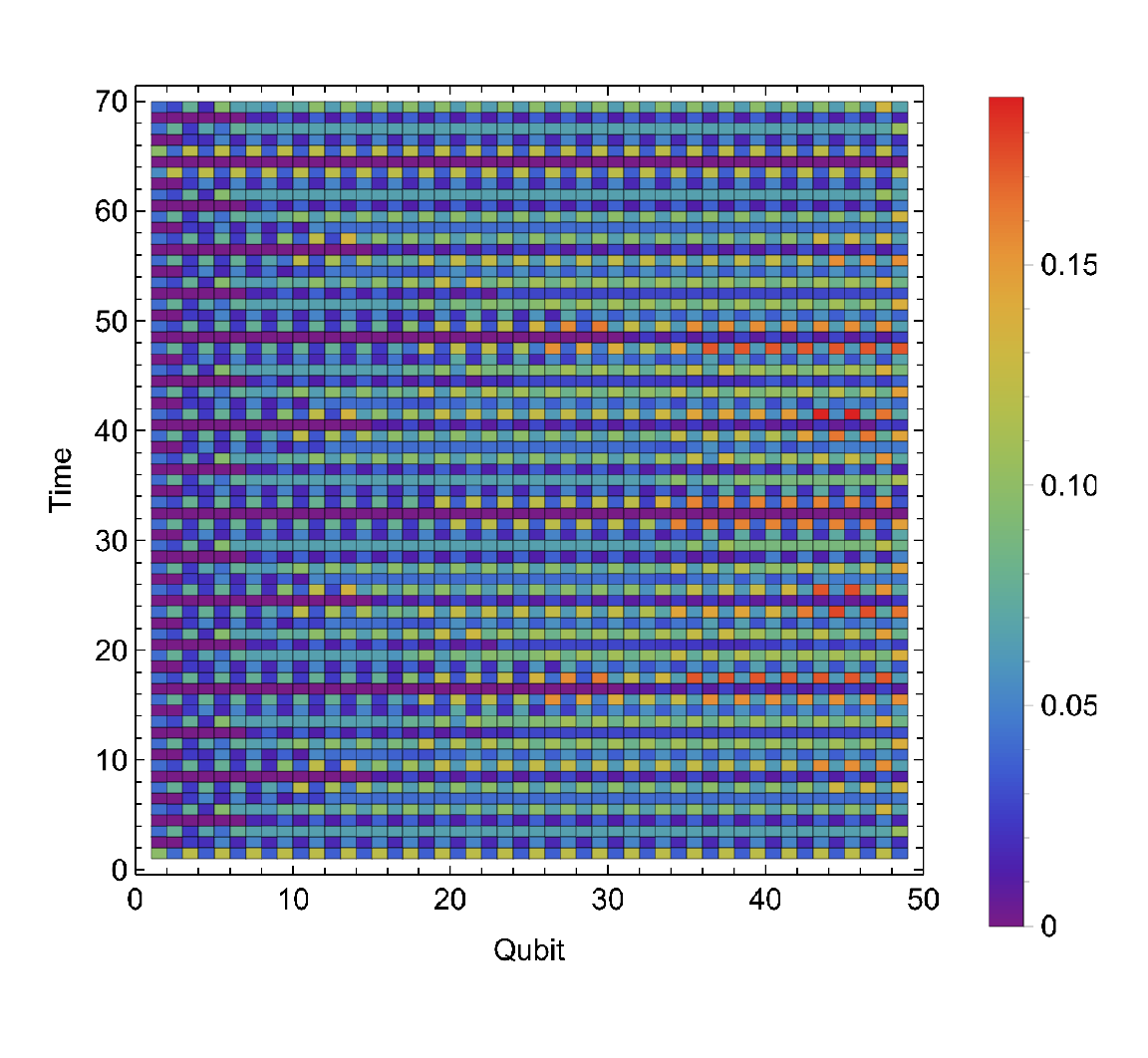}
    \caption{Mutual information between nearest neighbors for the left configuration in figure \ref{fig:ent1q_open} }
    \label{fig:MI_openZ}
\end{figure}

It is convenient for us to compute the mutual information between nearest neighbors for various setups. This is a measure of local correlations.
The mutual information is $M(i)= S(i)+S(i+1)-S(i,i+1)$, where the $S$ are the Von Neuman entropies of the single qubits and a pair of qubits,  These are evaluated at time $t$. We get the figures depicted in  figure \ref{fig:MI_openY} and  figure \ref{fig:MI_openZ}.
What we see from the figures is an alternating pattern of entanglement between nearest pairs of qubits. This states that how we pair qubots in the
time evolution at the beginning affects the entanglement entropy between the qubits. We also see in the left figure \ref{fig:MI_openY} that the entropy of pairs of nearest neighbors is maximal except at the first time before and after the periodicity of the system and the central slit.
There are essentially no short range correlations for most times. For the case of the states near the $Z$ eigenstate, from the pattern of mutual information in entanglement one still sees similarities with the single qubit information: there are approximately periodic patterns of entanglement that reflect local oscillations in periods of powers of two.  We also see that the mutual information between nearest neighbors is  not too large  almost always.

Notice that for the Y eigenstate case, all expectation values of the subalgebras either on $Z$ on the left, or the $X$ on the right vanish exactly, In these cases, the highest entropy state compatible with these local expectation values vanishing is maxinmally mixed. We expect this to be the endpoint of local thermalization in the thermodynamic limit.

\subsection{Periodic boundary conditions}

\begin{figure}
    \centering
    \includegraphics[width=3.2in] {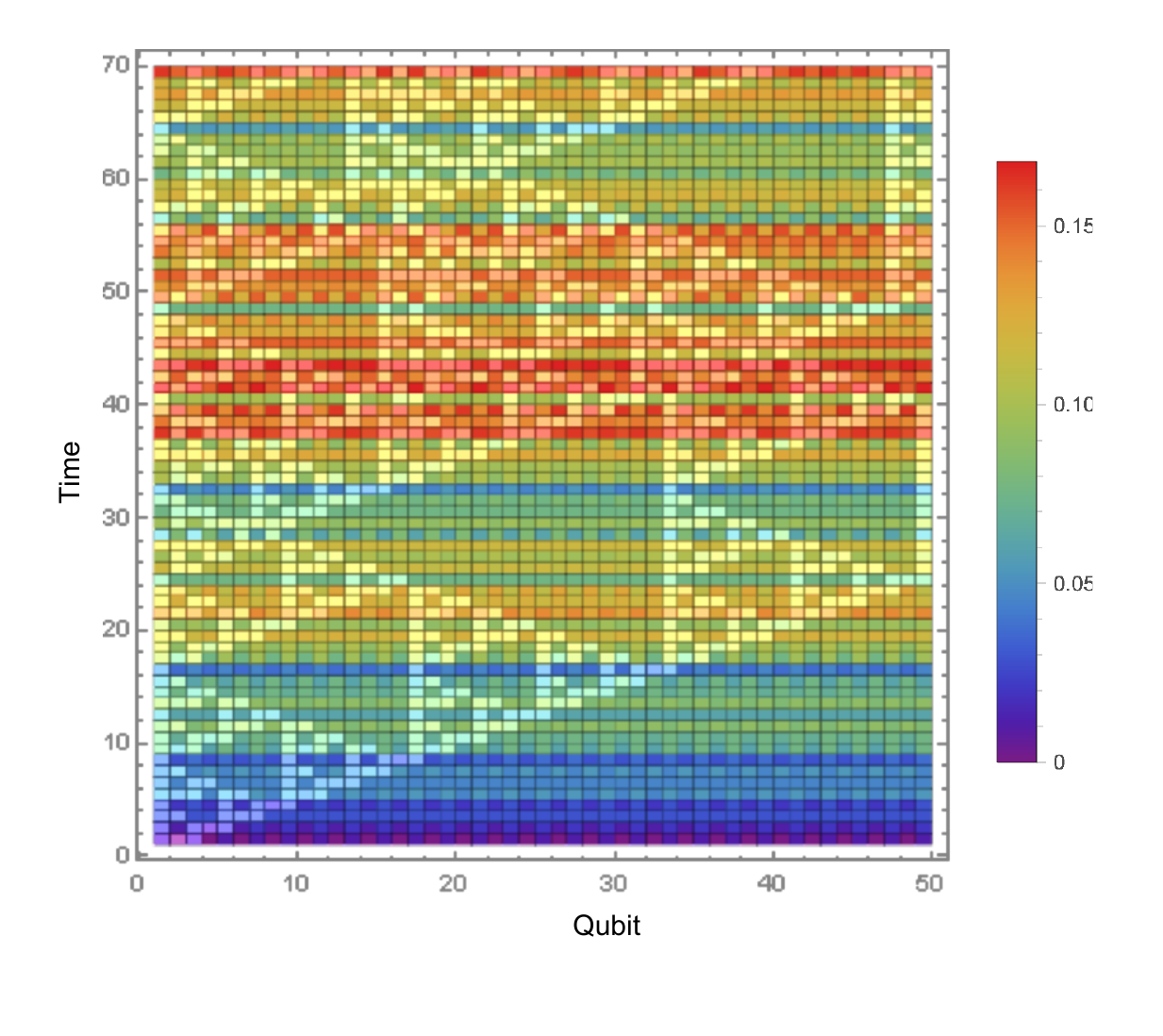}\includegraphics[width=3.2 in]{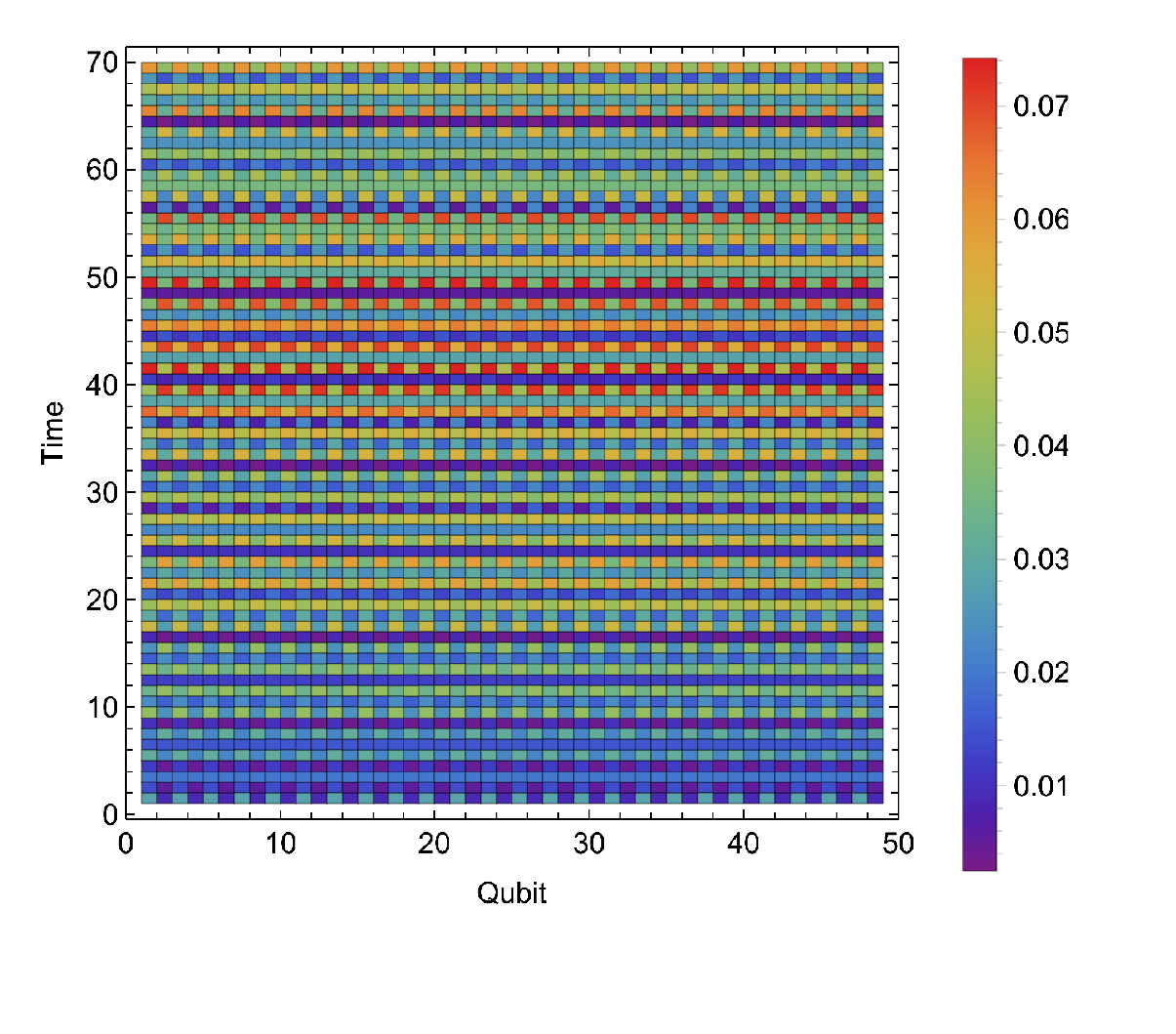}
    \caption{Example of periodic boundary conditions. Single quibt information on the left, for a state $\ket{\psi_0}\sim 0.734\ket0 +0.678\ket 1$, where we also show the wrapping of the Sierpinski triangle around the torus. The single qubit entropy itself is periodic modulo $2$ horizontally.
    Mutual information between nearest neighbors on the right. }
    \label{fig:periodic}
\end{figure}

The patterns of entanglement for closed boundary conditions are less interesting than for open boundary conditions. After all, they are locally periodic with period 2.   These are depicted in figure \ref{fig:periodic}.
We see that the single qubit entropy is large when there are a lot of images in the wrapped Sierpisnki triangle, The mutual information shows a spatial periodicity modulo $2$, as expected from the ordering of the quantum gate configurations. Also local mutual information is higher right before dips in the single qubit entropy show up, when the number of images in the wrapped Sierpinski triangle  is cur drastically from one step to the next (for example in the powers of two for short times, and near $t=40$.). These are consistent with the memory flashback discussion in section \ref{sec:Sierpinski}. The wrapping makes the periodicity long 
For a $Y$ eigenstate the local entropy is maximal after one step, and the mutual information vanishes after the first step, until the time periodicity can be seen. These are very similar to the early times of figure \ref{fig:MI_openY}, except that the maximally entangled region lasts
 a lot longer time.

\section{Conclusion}

We have studied a simple dynamical system where quantum equilibration can be understood analytically for a sufficiently simple class of initial states. The dynamics we have is closely related to the rule (60) cellular automaton, which produces the Sierpinski triangle fractal, but we get different slices of the fractal at fixed time.
In the thermodynamic limit the system equilibrates and approaches equilibrium subexponentially for a single quibit as follows
\begin{equation}
    \rho=\frac 12 + O(\exp(-a t^{h-1})
\end{equation}
where $h= \log_2(3)$ is the fractal dimension of the Sierpinski triangle. The result is not uniform in time, and moreover,  $a$ is a state dependent quantity that depends on the initial condition.
This should be compared to other thermalization situations where
\begin{equation}
    \rho=\frac 12 + O(\exp(- t/\tau^*)
\end{equation}
where $\tau^*$ is a decay time that is usually temperature dependent. By naive dimensional analysis, in a strongly coupled system the order of magnitude of $\tau^*$ is controlled by the temperature $T\sim 1/\beta$
of the final state. Here, we get slower convergence, controlled by the fractal dimension of the Sierpinski triangle, but where there is still some dependence on the initial state. This can be used to justify some approximate notion of temperature. 
The arguments we used suggest that this behavior is generic for bigger regions than one qubit.
In usual situations, a notion of temperature is associated to conservation of energy. Here, it is less clear if there is a local notion of conservation rule that plays a role. In open boundary conditions we found additional non-local conserved quantities that might play a similar role to  conservation of energy in giving an proxy for temperature: a conserved quantity that can predicts the $a$ value.

Here we have studied the late time equilibration dynamics of the system. It is also interesting to analyze the short time equilibration dynamics of large regions and their complement. Generically one expects linear growth in entanglement over time \cite{PhysRevX.8.021013,PhysRevX.8.021014}, somewhat independent of the initial product state. However, we have seen that states that are very close to the computational basis in the initial state stay that way for a while, this should delay the production of entropy before reaching saturation. Once near saturation,  the result we have discussed here would become important. To address that one would need to compute explicitly the expectation values of all Pauli strings in a region to get the reduced density matrix, similar to what we did with two sites. A big question is if this is transitory and then it is fixed  at some maximal value independent of the state, or if one reaches saturation before that happens.

There were special times, exponentially far apart, where log-quasi periodic deviations away from $\rho\sim 1/2$ are observed and which remember the initial condition, especially if the state is close to the computation basis, namely close to the $Z$ eigenstate at each site, or close to the $X$ eigenstate.

There was also two special states where $\rho=1/2$ exactly after a single step for all times. Namely, where we start with any of the two $Y$ eigenstates. These states equilibrate instantaneously at one site and probably saturate entanglement production between two large regions. It is interesting to study this further. It would be interesting to understand if this is a generic feature of the fractal 
clifford cellular automata. It is also interesting to understand more generally how top compute their fractal dimensions: these control the approach to equilibration.

It is obviously interesting to understand to what extent one can get similar behaviors for other such cellular automaton quantum dynamical systems. This problem has been partially analyzed in \cite{schlingemann2008structure,gutschow2010time} under restricted conditions of periodicity in the lattice dynamics. It is likely that their results, properly understood, hold generally in other  systems  like we have studied and that there is a generalization of their methods and their classification of automata that would apply to all these cases.

If the classification holds in general, we expect that systems with gliders will never thermalize the 
glider observables. Also the periodic ones will not thermalize. Only the ones that are fractal (for all Pauli operators) would do so and our arguments suggest that the approach to equilibrium is subexponential (this is for one dimensional systems). Needless to say, one would need to analyze more general initial conditions to get a good picture of general equilibration phenomena in these settings.

This would hopefully not only dictate when fractals are produced, but also what dimensions they might have.  Situations with gliders,   should  be very similar to a similar automata as above where one iterates  swap gate between layers. There, each classical qubit is a glider.
In that system, entanglement between the two sides was in a certain sense maximal for all times for sufficiently generic initial states where two halves are disconnected initially (see \cite{Berenstein:2019axm}). We expect the same to hold in that case.

We are also interested in implementing this dynamics in actual quantum devices and to check the results of the quantum computation to the theory prediction. In particular the log periodic memory and the oscillations near the boundary case suggest that one can use this property to track how the real NISQ system introduces errors and deteriorates in time, where we have simple predictions for the density matrix at each site.

\acknowledgments
D.B. Would like to thank R. Brower, J. Broz,  Y. Meurice, S. Mukherje, M. Srednicki, S. Vijay for discussions. 
Work of D. B. supported by the Department of Energy grant DE-SC0019139.


\begin{thebibliography}{99}

\bibitem{cazalilla2010focus}
M.~A. Cazalilla and M.~Rigol
``Focus on dynamics and thermalization in isolated quantum many-body systems", New Journal of Physics,
vol 12, no 5, pp. 055006, (2010)

\bibitem{arute2019quantum}
Arute, Frank and Arya, Kunal and Babbush, Ryan and Bacon, Dave and Bardin, Joseph C and Barends, Rami and Biswas, Rupak and Boixo, Sergio and Brandao, Fernando GSL and Buell, David A and others, ``Quantum supremacy using a programmable superconducting processor",
Nature, \textbf{ 574}, pp 505--510, (2019) 

\bibitem{gopalakrishnan2018facilitated}
Gopalakrishnan, Sarang and Zakirov, Bahti, 
``Facilitated quantum cellular automata as simple models with non-thermal eigenstates and dynamics", Quantum Science and Technology, Vol 3, no.4 , p. 044004, 2018  [arxiv:1802.07729]


\bibitem{deutsch1991quantum}
J.~Deutsch 
``Quantum statistical mechanics in a closed system"
Phys. Rev. A. 43, 2046 (1991)

\bibitem{PhysRevE.50.888}
M.~ Srednicki, 
``Chaos and quantum thermalization",Phys. Rev. E 50, 2, p 888-991, 1994
doi: {10.1103/PhysRevE.50.888}





\bibitem{Berenstein:2018zif}
D.~Berenstein,
``A toy model for time evolving QFT on a lattice with controllable chaos,''
[arXiv:1803.02396 [hep-th]].

\bibitem{Berenstein:2015yxu}
D.~Berenstein and A.~M.~Garcia-Garcia,
``Universal quantum constraints on the butterfly effect,''
[arXiv:1510.08870 [hep-th]].

\bibitem{alba2019operator}
Alba, Vincenzo and Dubail, Jerome and Medenjak, Marko,
"Operator entanglement in interacting integrable quantum systems: the case of the rule 54 chain"
Physical review letters, 122, 25, 250603, 2019



\bibitem{Gottesman:1998hu}
D.~Gottesman,
``The Heisenberg representation of quantum computers,'', Published in: Group22: Proceedings of the XXII International Colloquium on Group Theoretical Methods in Physics, eds. S. P. Corney, R. Delbourgo, and P. D. Jarvis, pp. 32-43 (Cambridge, MA, International Press, 1999), p.32-43
[arXiv:quant-ph/9807006 [quant-ph]].

\bibitem{schlingemann2008structure}
Schlingemann, Dirk-M and Vogts, Holger and Werner, Reinhard F,
``On the structure of Clifford quantum cellular automata", 
Journal of Mathematical Physics, vol 49,  no 11, 112104, 2008

\bibitem{gutschow2010time}
G{\"u}tschow, Johannes and Uphoff, Sonja and Werner, Reinhard F and Zimbor{\'a}s, Zolt{\'a}n, ``Time asymptotics and entanglement generation of Clifford quantum cellular automata"
Journal of mathematical physics, vol 51, no 1, 015203, 2010. 


\bibitem{Farrelly:2019zds}
T.~Farrelly,
``A review of Quantum Cellular Automata,''
Quantum \textbf{4}, 368 (2020)
doi:10.22331/q-2020-11-30-368
[arXiv:1904.13318 [quant-ph]].

\bibitem{s41467-019-09757-y}
Bu\v{c}a, B., Tindall, J. , Jaksch, D. ``Non-stationary coherent quantum many-body dynamics through dissipation." Nat Commun 10, 1730 (2019). https://doi.org/10.1038/s41467-019-09757-y


\bibitem{PhysRevB.102.041117}
Medenjak, Marko and Bu\v{c}a, Berislav and Jaksch, Dieter,
``Isolated Heisenberg magnet as a quantum time crystal", Phys. Rev. B 102, vol 4, 041117, 
 doi = {10.1103/PhysRevB.102.041117}
 
\bibitem{PhysRevX.8.021013}
von Keyserlingk, C. W. and Rakovszky, Tibor and Pollmann, Frank and Sondhi, S. L., ``Operator Hydrodynamics, OTOCs, and Entanglement Growth in Systems without Conservation Laws", Phys. Rev. X, vol 8, no 2., 021013, 2018 
 
\bibitem{PhysRevLett.125.060601} 
 Chinzei, Koki and Ikeda, Tatsuhiko N., ``Time Crystals Protected by Floquet Dynamical Symmetry in Hubbard Models", Phys. Rev. Lett. {\bf 125}, 060601
 
 
\bibitem{Berenstein:2019axm}
D.~Berenstein and D.~Teixeira,
``Maximally entangling states and dynamics in one dimensional nearest neighbor Floquet systems,''
[arXiv:1901.02944 [hep-th]].




\bibitem{PhysRevX.8.021014}
Nahum, Adam and Vijay, Sagar and Haah, Jeongwan, ``Operator Spreading in Random Unitary Circuits", Phys. Rev. X, vol 2, 021014, 
2018.





\end{thebibliography}
\end{document}